\documentclass[12pt]{article}
\input epsf
\makeatletter

\def\newpic#1{}
\def\hybrid{\topmargin 0pt      \oddsidemargin 0pt
                    \headheight 0pt \headsep 0pt

                  \textwidth 6.25in       % A4 paper
                    \textheight 9.5in       % A4 paper
                    \marginparwidth 0.0in
                    \parskip 5pt plus 1pt   \jot = 1.5ex}
\catcode`\@=11
\def\marginnote#1{}

\newcount\hour
\newcount\minute
\newtoks\amorpm
\hour=\time\divide\hour by60
\minute=\time{\multiply\hour by60 \global\advance\minute by-\hour}
\edef\standardtime{{\ifnum\hour<12 \global\amorpm={am}%
                    \else\global\amorpm={pm}\advance\hour by-12 \fi
                    \ifnum\hour=0 \hour=12 \fi
                    \number\hour:\ifnum\minute<10
0\fi\number\minute\the\amorpm}}
\edef\militarytime{\number\hour:\ifnum\minute<10 0\fi\number\minute}

\def\draftlabel#1{{\@bsphack\if@filesw {\let\thepage\relax
               \xdef\@gtempa{\write\@auxout{\string
                  \newlabel{#1}{{\@currentlabel}{\thepage}}}}}\@gtempa
               \if@nobreak \ifvmode\nobreak\fi\fi\fi\@esphack}
                    \gdef\@eqnlabel{#1}}
\def\@eqnlabel{}
\def\@vacuum{}
\def\draftmarginnote#1{\marginpar{\raggedright\scriptsize\tt#1}}

\def\draftlabel#1{{\@bsphack\if@filesw {\let\thepage\relax
               \xdef\@gtempa{\write\@auxout{\string
                  \newlabel{#1}{{\@currentlabel}{\thepage}}}}}\@gtempa
               \if@nobreak \ifvmode\nobreak\fi\fi\fi\@esphack}
                    \gdef\@eqnlabel{#1}}
\def\@eqnlabel{}
\def\@vacuum{}
\def\draftmarginnote#1{\marginpar{\raggedright\scriptsize\tt#1}}

\def\draft{\oddsidemargin -.5truein
                    \def\@oddfoot{\sl preliminary draft \hfil
                    \rm\thepage\hfil\sl\today\quad\militarytime}
                    \let\@evenfoot\@oddfoot \overfullrule 3pt
                    \let\label=\draftlabel
                    \let\marginnote=\draftmarginnote

\def\@eqnnum{(\theequation)\rlap{\kern\marginparsep\tt\@eqnlabel}%
\global\let\@eqnlabel\@vacuum}  }

\def\numberbysection{\@addtoreset{equation}{section}
                    \def\theequation{\thesection.\arabic{equation}}}

\def\underline#1{\relax\ifmmode\@@underline#1\else
                    $\@@underline{\hbox{#1}}$\relax\fi}

\def\titlepage{\@restonecolfalse\if@twocolumn\@restonecoltrue\onecolumn
                 \else \newpage \fi \thispagestyle{empty}\c@page\z@
                    \def\thefootnote{\fnsymbol{footnote}} }

\def\endtitlepage{\if@restonecol\twocolumn \else  \fi
                    \def\thefootnote{\arabic{footnote}}
                    \setcounter{footnote}{0}}  %\c@footnote\z@ }
\relax

\makeatletter
\newdimen\normalarrayskip              % skip between lines
\newdimen\minarrayskip                 % minimal skip between lines
\normalarrayskip\baselineskip
\minarrayskip\jot
\newif\ifold             \oldtrue            \def\new{\oldfalse}
\def\arraymode{\ifold\relax\else\displaystyle\fi} % mode of array entries
\def\eqnumphantom{\phantom{(\theequation)}}     % right phantom in eqnarray
\def\@arrayskip{\ifold\baselineskip\z@\lineskip\z@
                \else
                \baselineskip\minarrayskip\lineskip2\minarrayskip\fi}
\def\@arrayclassz{\ifcase \@lastchclass \@acolampacol \or
\@ampacol \or \or \or \@addamp \or
              \@acolampacol \or \@firstampfalse \@acol \fi
\edef\@preamble{\@preamble
             \ifcase \@chnum
                \hfil$\relax\arraymode\@sharp$\hfil
                \or $\relax\arraymode\@sharp$\hfil
                \or \hfil$\relax\arraymode\@sharp$\fi}}
\def\@array[#1]#2{\setbox\@arstrutbox=\hbox{\vrule
                height\arraystretch \ht\strutbox
                depth\arraystretch \dp\strutbox
                width\z@}\@mkpream{#2}\edef\@preamble{\halign
\noexpand\@halignto
\bgroup \tabskip\z@ \@arstrut \@preamble \tabskip\z@ \cr}%
\let\@startpbox\@@startpbox \let\@endpbox\@@endpbox
             \if #1t\vtop \else \if#1b\vbox \else \vcenter \fi\fi
             \bgroup \let\par\relax
             \let\@sharp##\let\protect\relax
             \@arrayskip\@preamble}
\def\eqnarray{\stepcounter{equation}%
                         \let\@currentlabel=\theequation
                         \global\@eqnswtrue
                         \global\@eqcnt\z@
                         \tabskip\@centering
                         \let\\=\@eqncr
            \halign to \displaywidth\bgroup
               \eqnumphantom\@eqnsel\hskip\@centering
               $\displaystyle \tabskip\z@ {##}$%
               \global\@eqcnt\@ne \hskip 2\arraycolsep
                    $\displaystyle\arraymode{##}$\hfil
               \global\@eqcnt\tw@ \hskip 2\arraycolsep
                    $\displaystyle\tabskip\z@{##}$\hfil
                    \tabskip\@centering
               &{##}\tabskip\z@\cr}
\begingroup\ifx\undefined\newsymbol \else\def\input#1 {\endgroup}\fi
\newfont{\hr}{msbm10}
\newfont{\ams}{msam10}
\def\beq{\begin{equation}}
\def\eeq{\end{equation}}
\def\ba{\beq\new\begin{array}{c}}
\def\ea{\end{array}\eeq}

\hybrid

\def\beq{\begin{equation}}
\def\eeq{\end{equation}}
\def\p{\partial}

\begin{document}

\begin{titlepage}

\title{Large scale correlations in normal  non-Hermitian
matrix ensembles}

\author{P. Wiegmann \thanks{James Frank Institute and Enrico Fermi
Institute
of the University of Chicago, 5640 S.Ellis Avenue,
Chicago, IL 60637, USA and
Landau Institute for Theoretical Physics, Moscow, Russia}
\and A. Zabrodin
\thanks{Institute of Biochemical Physics,
Kosygina str. 4, 119991 Moscow, Russia
and ITEP, Bol. Cheremushkinskaya str. 25, 117259 Moscow, Russia}}

\date{October 2002}
\maketitle
%\vspace{-7cm}

%\centerline{
%\hfill ITEP/TH-32/01}
%\centerline{
%\hfill FIAN/TD-11/01}

%\vspace{7cm}

\begin{abstract}

We compute the large scale (macroscopic) correlations in
ensembles of normal random matrices
with a general non-Gaussian  measure and in ensembles of general
non-Hermition matrices with a  class of non-Gaussian measures. In both
cases the eigenvalues are complex  and in the large $N$
limit they occupy a domain in the
complex plane.  For the case when the support of eigenvalues is a connected
compact domain, we compute two-, three-  and  four-point connected
correlation functions in the first non-vanishing order in $1/N$ in a manner
that the algorithm of computing higher correlations becomes clear. The
correlation functions are expressed through the solution of the Dirichlet
boundary problem in the domain
complementary to the support of eigenvalues.

\end{abstract}

\vfill
ITEP-TH-49/02

\end{titlepage}

\section{Introduction}

A matrix $M$ is called normal if it commutes with its
Hermitian conjugated: $[M, M^{\dag}]=0$.
The both matrices can be simultaneously diagonalized,
the eigenvalues being complex numbers.
The partition function of normal matrices has the
general form
\beq\label{ZN}
Z_N =\int_{\rm normal} d\mu (M)
e^{\frac{1}{\hbar}{\rm tr} \, W(M, M^{\dag})}.
\eeq
Here $\hbar$ is a parameter, and the measure of integration
over normal $N\times N$ matrices
is induced from the flat metric on the space
of all complex matrices.
Like in other random matrix models,
the large $N$ limit of interest implies
$N\to \infty$, $\hbar \to 0$, while $N\hbar$ stays finite.

In a particular case, when Laplacian of the potential
$W(z,\bar z)$ is a constant
in a big domain of a complex plane, i.e.,
\beq\label{W}
W=-MM^{\dag}+V(M)+\bar V(M^{\dag}),
\eeq
where $V(z)$ and $\bar
V(z)$ are holomorphic functions,
the normal matrix  ensemble is equivalent to the ensemble
of general complex matrices. It  generalizes
the Ginibre-Girko Gaussian ensemble \cite{Ginibre}.
In this, perhaps the most
interesting case for applications, the model bears some formal
similarities with the model of two Hermitian random matrices
\cite{DKK} and the matrix quantum mechanics in a singlet sector
\cite{c1,MQM}.
Unlike models of few Hermitian matrices,
the normal matrix model  is integrable for a general class of potentials,
not only of the form (\ref{W}).

Applications and studies of matrix ensembles with complex eigenvalues are
numerous. A large list of references can be found in recent papers
\cite{recent}. New applications
to diffusion limited growth models (Laplacian growth) \cite{MWWZ},
complex analysis \cite{WZ}-\cite{MWZ}
and Quantum Hall effect \cite{ABWZ} were found recently.

Despite of a comprehensive literature  on the
model of one and two Hermitian matrices, interest to
the normal matrix ensemble, first introduced
in \cite{ChauYu} and further studied
in \cite{ZC}, as well as to a model of general complex matrices with
potential (\ref{W}), just starts to grow.  In
this paper we revisit
the ensemble of normal random matrices to calculate  the large
$N$ limit of correlation functions under
condition that separation between arguments
is much more than an average distance
between eigenvalues (macroscopic, or smoothed
correlations). Short scale
(microscopic) correlations in the ensemble with the potential
(\ref{W}) are well studied (see e.g. \cite{F} for a review).

At large $N$ the eigenvalues of the random matrices are distributed
within a domain
(with sharp edges) of a complex plane with a density proportional to
$\hbar^{-1}\Delta W$,
where $\Delta$ is the 2D Laplace operator.
We assume that
       the support of
eigenvalues is a single connected bounded domain $D$, and that the
boundary is a Jordan curve.
Analytical properties of this curve determines the correlation functions.

We will show that  correlation functions
are expressed through
the objects of  the Dirichlet boundary problem of
the domain complimentary to the
support of eigenvalues. Namely, the two-point correlations are expressed
through the
Dirichlet Green function, while  higher multi-point correlations  are
expressed through the Neumann jump on the boundary,
through the Bergman kernel and through the curvature of the boundary.
The objects similar to the correlation functions
previously appeared in studies of thermal  fluctuations in classical
confined Coulomb plasma \cite{J} (see also \cite{F} for a review) and in
recent studies of integrable structure of the Dirichlet boundary problem
in \cite{MWZ}.
If the potential $W$ is such that the support of eigenvalues
collapses to a cut (or cuts), the normal matrix
model reproduces known large $N$-limit features of the Hermitian matrix
model.

We compute two-, three- and four-point connected functions of density
of eigenvalues in the
leading order in $N$, in a manner that the algorithm of computing higher
correlation functions becomes clear.
Connected density correlators
are localized at the edge of the support
of eigenvalues and show some universal features. They depend on the
potential only through a shape of the  support of the eigenvalues and
     boundary values of  a
finite number of  derivatives of the potential.
In the literature on Hermitian random matrices the universal character of
     correlations has been emphasized in Refs.
\cite{univ1,univ2}. The
two-point functions are distinguished since they depends on the
support of the eigenvalues only. The two-point function has been
previously computed in Refs. \cite{J} in the course of
studies of confined Coulomb plasma.

Other aspects, including
integrable structure of the model
(discussed already in \cite{ZC,int}),
semiclassical properties of  biorthogonal polynomials,
critical points (a boundary with cusps),
disconnected supports of eigenvalues,
correlations at short distances, etc.,
will be addressed elsewhere.

\section{Preliminaries}

\paragraph{The measure of normal matrices.}
The measure in (\ref{ZN}) is induced from the flat
metric $|| \delta M ||^2 =
\mbox{tr}\, (\delta M \delta M^{\dag})$
in the space of all complex matrices. Formally,
one can write $d\mu (M)=\delta ([M, M^{\dag}])dM dM^{\dag}$,
where the delta-function selects the subspace of
normal matrices.
To introduce coordinates in this subspace,
one  uses of the decomposition
$M=UZU^{\dag}$ , where
$U$ is a unitary matrix and $Z=\mbox{diag}\,(z_1,
\ldots , z_N )$ is a diagonal matrix of eigenvalues
.
The measure is then given by
\beq\label{E1}
d\mu (M)=\frac{d\mu_{0}(U)}{N! \,\mbox{Vol}\, ({\cal U}(N))}
|\Delta_{N}(z)|^2 \prod_{i=1}^{N}d^2z_i,
\eeq
here $d^2z \equiv dx dy$ for $z=x+iy$, $d\mu_0$ is
the Haar measure on the unitary group ${\cal U}(N)$, and
$\Delta_N(z)=\det (z_{j}^{i-1})_{1\leq i,j\leq N}=
\prod_{i>j}^{N}(z_i -z_j)$
is the Vandermonde determinant.

If  $A(M)$ is any invariant function
(i.e., a symmetric function of
eigenvalues) of a  matrix
          $M$ (and $M^{\dag}$), then
the mean value
\beq\label{ZN3}
\left < A \right >
=\frac{\int d\mu (M) A(M)e^{\frac{1}{\hbar}W}}
{\int d\mu (M) e^{\frac{1}{\hbar}W}}
\eeq
is expressed through the integral over eigenvalues
\beq\label{mean}
\left <A \right > =\frac{1}{N! Z_N} \int A(z )
|\Delta_N (z)|^2 \prod_{j=1}^{N} \left (
e^{\frac{1}{\hbar}W(z_j , \bar z_j )}d^2 z_j \right ),
\eeq
where the partition function is\beq\label{E2}
Z_N = \frac{1}{N!}\int |\Delta_{N}(z)|^2
\prod_{j=1}^{N}\left ( e^{\frac{1}{\hbar}W(z_j , \bar z_j )}
d^2z_j \right ).
\eeq

\paragraph{Potential.}
For notational simplicity, we shall write simply
$W(z)$ instead of $W(z, \bar z)$.
We assume that $W$ is a real-valued function
with (at least local) minimum at the origin and
set $W(0)=0$ for convenience. We also assume that
the integral (\ref{E2}) converges and that $W$ is a regular function
in both variables
at the origin. We shall also set
\beq\label{sigma}
\sigma (z)=-\frac{1}{\pi}\p_z \p_{\bar z}W(z)=-\frac{1}{4\pi}\Delta W,
\eeq
and assume  that
$\sigma (z) >0$.

A special interesting case \cite{ABWZ} occurs if the potential is harmonic
in some big domain around the origin. Then in this domain
$\sigma(z)=\mbox{constant}$, say, set to be $\pi^{-1}$, and
\beq\label{potential}W=-|z|^2+V(z)+\bar
V(\bar z),
\eeq
where $V(z)$ is a holomorphic function.

\paragraph{The measure of complex matrices.}If the potential is chosen in
the form
(\ref{potential}), then
the same measure, up to a numerical factor,
appears in complex random matrices.
In this case the relevant decomposition reads: $M=U(Z+R)U^\dagger$,
where $U$, $Z$
are again unitary and diagonal matrix  respectively, and $R$
is an upper triangular matrix.
The measure (\ref{E1}) acquires a
multiplicative  factor
$\prod_{ij}d R_{ij}e^{-R_{ij}^2}$ (we use the fact that
$\mbox{tr} \,M M^\dag=
\mbox{tr}\, ZZ^\dag +
\mbox{tr}\, RR^\dag$). As  a result the representation (\ref{mean})
holds \cite{Mehta}.

\paragraph{Coherent states of particles in magnetic field.}
Most of the known matrix ensembles   represent  coherent
states of $N$ fermions. Fermonic representations are well
known in the case
of Hermitian matrices (see, e.g., \cite{fermions}), where fermions live
on  a line and are confined by a potential.

Complex matrix ensembles also enjoy a
fermionic representation \cite{FGIL,ABWZ}.
In this case fermions are situated on a plane and occupy the lowest
energy level of a strong magnetic field $B(z)$,
the magnetic field is not necessarily
uniform. It is related
to the potential by
$B(z)=2\pi \sigma(z)$. The coherent state of
the fully occupied lowest level is
\beq\label{psi}
\Psi(z_1,\dots,z_N)=
\frac{1}{\sqrt{N!}}\Delta_N (z) \,e^{\sum_{j=1}^{N}
\frac{1}{2\hbar}W(z_j)},\quad Z_N=\int |\Psi(z_1,\dots,z_N)|^2 \prod d^2
z_i.
\eeq
So $Z_N$, in this picture, is the normalization factor of
the wave function.

\paragraph{Example: Gaussian model.}
In the case
$$
W(z)=-\pi\sigma |z|^2 +2 {\cal R}e (t_1 z +t_2 z^2 )\,,
\;\;\;\;\; \sigma >0\,,
$$
(the Ginibre-Girko ensemble)
the partition function can be found
explicitly even for finite $N$
(see e.g. \cite{FGIL}):
\beq\label{gaus1}
Z_N (\sigma, t_1 , t_2 ) = Z_{N}^{(0)}
(\pi\sigma)^{\frac{1}{2}(N^2 -N)}
(\pi^2\sigma^2 -4t_2 \bar t_2 )^{-\frac{1}{2}N^2}
\exp \left ( \frac{N}{\hbar}\,
\frac{t_{1}^{2}\bar t_2 +\bar t_{1}^{2}
t_2 + \pi \sigma |t_1|^2}{\pi^2 \sigma^2 -4t_2 \bar t_2}\right ),
\eeq
where $Z_{N}^{(0)}=
\hbar^{\frac{1}{2}N(N+1)}\pi^N
\prod_{j=1}^{N-1} j!$ is the partition function
for $W=-|z|^2$.
Correlation functions in this case are expressed through Hermite
polynomials and are known explicitly for any $N$.

\paragraph{Correlation functions of traces.}
In this paper, we are interested in correlation functions
of products of $n$ traces,
$\left <\prod_{i=1}^{n}
\, \mbox{tr} \, f_i (M)\right >$.
Clearly, they
are expressed through $n$-point correlation functions of
the density of eigenvalues,
$$
\rho (z) =\hbar\sum_{i=1}^{N}\delta^{(2)}(z-z_i ).
$$
We obviously have:
\beq\label{mean1}
\hbar^n\left <\prod_{i=1}^{n}\,
\mbox{tr} \, f_i (M)\right >=
\int \left <\rho (z_1) \ldots \rho (z_n)
\right > f_1 (z_1  )
\ldots f_n (z_n ) \prod_{j=1}^{n} d^2 z_j.
\eeq
So, the density correlation functions carry the necessary  information.

It is customary to deal with connected part
of a correlation function. In case of the 2-point
function it is
$$
\left < \rho (z_1)\rho (z_2)\right >_{\rm conn}=
\left < \rho (z_1) \rho (z_2)\right >-
\left < \rho (z_1)\right >
\left < \rho (z_2)\right >.
$$
As $N\to \infty$, the $n$-point correlation
function of densities is $O(1)$
while the connected part of the $n$-point function
is $O(N^{2-2n})$.

\paragraph{The method of functional derivatives.}
Correlation functions are the linear responses to a small variation of
the potential.  Variation of the partition function
(\ref{E2}) over a general potential $W(z)$ inserts
$\frac{1}{\hbar}\sum_i \delta^{(2)}(z-z_i)$ into the integral. Then
\beq\label{FD2}
\left <\rho (z)\right >=
\hbar^2  \,\frac{\delta \log Z_N}{\delta W(z)}\,,
\;\;\;\;
\left <\rho (z_1) \rho (z_2) \right >_{{\rm conn}}=\hbar^2
\frac{\delta \left <\rho (z_1)\right >}{\delta W(z_2)}\,,
\eeq
and, more generally,
$$
\left < \rho (z_1)\ldots \,\rho (z_n)\right >_{{\rm conn}}=
\hbar^{2n}\,\frac{\delta^n
\log Z_N }{\delta W(z_1)\ldots \delta W(z_n)}.
$$

We shall use this method in
the following version. Set $\delta W(z) =\varepsilon g(z)$,
where $g$ is an arbitrary smooth function on the plane and
$\varepsilon \to 0$. Then, in the first order in $\varepsilon$,
\beq\label{FD5}
\hbar^2 \delta \left < \mbox{tr}\, f(M)\right >
=\varepsilon
\left < \mbox{tr}\, f(M)
\, \mbox{tr}\, g(M) \right >_{{\rm conn}}.
\eeq

It is often convenient to
consider correlations of the potential
\beq\label{bose}
\varphi(z)=\int\log|z-z'|^2\rho(z')d^2z'=\hbar\mbox{ tr}\left(
\log(z-M)(\bar z-M^\dag)\right),
\eeq
and of the current field
\beq\label{current}
J(z) \equiv\p\varphi(z)=\hbar
\, \mbox{tr}\left(\frac{1}{z-M}\right),
\eeq
rather than correlations of density.
The potential is the  Bose field or the loop field of
a (collective) field theory of the matrix model
(more
accurately, $\varphi$ is a negative mode
part of a Bose field). The field theory is proved to be successful
approach to Hermitian matrix ensembles \cite{Jev}.
Elsewhere we will develop this approach for the non-Hermitian ensembles.
The potential  is harmonic
outside the support of eigenvalues except at
infinity where it behaves as $2N\hbar\log|z|$.
The current is holomorphic outside the support of eigenvalues.

          In order to obtain the correlations of the potentials, one
has to vary the partition  function by $W(z)\to W(z)+
\varepsilon\log|z-\zeta|^2$
where $\zeta$ is a parameter.
We denote this particular deformation of the potential
by $\delta_\zeta$:
\beq\label{po}
\delta_\zeta W(z)=\varepsilon\log|z-\zeta|^2.
\eeq
Under this variation the
correlation function changes by insertion of the field $\varphi(\zeta)$:
\beq\label{FD6}
\hbar^2 \delta_\zeta \left < A\right >
=\varepsilon
\left < A
\, \varphi(\zeta) \right >_{{\rm conn}}.
\eeq
This is the linear responce relation used in the
Coulomb gas theory \cite{F}.

While varying the potential it is  important to distinguish a
harmonic variation of the
potential $W(z,\bar z)\to W(z,\bar z)+V(z)+\bar
V(\bar z)$, where $V(z)$ is a holomorphic function. This variation does
not change $\Delta W$.
To implement a harmonic  variation one may  extend the  potential by
adding  a harmonic
          function $W\to W+2{\cal R}e \,
\sum_{k\geq 1}t_kz^k$
and apply the operator
         $D(z) =\sum_{k\geq 1}\frac{z^{-k}}{k}\p_{t_k}$
used in Refrs.\,\cite{WZ}-\cite{MWZ}.
Then correlators of holomorphic parts of the potential
$\phi(z)=\hbar\mbox{tr}\log(z-M)$ are
\beq\label{FD4}\begin{array}{l}
\left < \phi(z)\right >=
\hbar N \log z
-\hbar^2 { D}(z) \log Z_N,
\\ \\
\displaystyle{
\left < \prod_{i=1}^{n}
\phi(z_i) \right >_{{\rm conn}}=
(-1)^n \hbar^{2n} \prod_{i=1}^{n}{ D}(z_i ) \log Z_N\,,
\;\;\;\;n\geq 2}.
\end{array}
\eeq

\paragraph{Dirichlet boundary problem.} We list some elements of the
external Dirichlet boundary problem, which
are extensively used below. More details can be found in
\cite{C-H,Gakhov}.

Let $D$ be a closed connected domain of the complex plane bounded
by a smooth curve. Given a real analytic
function $f(z)$ in a vicinity of the boundary, we may restrict it to the
boundary of $D$.
The external Dirichlet problem is to find
a harmonic function in the exterior of $D$, whose value on the
boundary is
$f(z)$. We call this
harmonic function the {\it harmonic extension} of $f(z)$ to the
exterior and denote it by $f^H(z)$.
It is given by the
\begin{itemize}
\item Dirichlet formula:
$$f^H(z)=-\frac{1}{2\pi}\oint\p_{n'}
G(z,z') f(z') |dz'|,$$
where $\p_{n'}$
is the normal derivative at the boundary with respect to the
second variable, with
the normal vector being directed to the exterior of the
domain $D$.

The Green function $G(z,z')$ for the exterior problem is a
harmonic function
everywhere outside $D$
including infinity except the point $z=z'$,
where it has a logarithmic singularity:
$G(z,z')\to \log |z-z'|$ as $z\to z'$.
If $z'\to\infty$, then $G(z, \infty)\to -\log |z|$.
The Green function is symmetric in $z,z'$ and
vanishes on the boundary.
In particular, the harmonic extension of
$\log |z-\zeta |$ in $z$ is
\beq\label{log}
          (\log|z-\zeta|)^H=
\log |z-\zeta | -
G(z,\zeta)+G(z,\infty).
\eeq

If the point $\zeta$ happens to be inside,
$G(z,\zeta)$ in this formula is understood to be null.
\item
The Green function can be expressed through a conformal map
$w(z)$ from ${\bf C}\setminus D$
(the exterior of $D$) onto the exterior of the  unit disk:
\beq\label{D}
G(z,z')=\log \left |\frac{w(z)-w(z')}{1-\overline{w(z)}w(z')}
\right |.
\eeq
This formula does not depend on the normalization of the map.
It is convenient to fix $w(z)$
by the condition that it sends infinity to infinity
and the coefficient in front of the leading term
as $z\to \infty$ is real positive.

\item{Neumann external jump} is the
difference between normal
derivatives of a smooth function at the boundary and its harmonic
extension. The Neumann external jump operator ${\sf R}$
acts as follows:
\beq\label{R}
f(z)\mapsto ({\sf R}f)(z)=
\p_{n}^{-}(f(z)\!-\!f^{H}(z)),\quad \quad z\in\p D.
\eeq
The upperscript indicates that the
derivative is taken in the exterior of the boundary.
As it follows from the Dirichlet formula,
the Neumann jump is an integral operator
on the boundary curve with the kernel
given by normal derivative of the Green function in both
arguments:
$({\sf R}f)(z)=\p_n f(z)
+\frac{1}{2\pi}\oint\p_n\p_{n'}G(z,z')f(z')|dz'|$.
In fact this integral is not yet well-defined
since the kernel has a second order pole
on the contour. The operations of taking
normal derivative and contour integration do not commute. The above
formula has to be understood as $\p_n\oint\p_{n'}G(z,z')f(z')|dz'|$
Alternatively
the formula is understood as
the  principal value integral:
\beq\label{CC51}
({\sf R}f)(z)=\p_n f(z)+
\frac{1}{2\pi}
\mbox{P.V.} \oint \p_n \p_{n'}G(z,z')
(f(z)-f(z' ))|dz'|.
\eeq

\item {Hadamard formula}
describes deformation of the Dirichlet Green function
under deformation of the domain.

\begin{figure}[tp]
\epsfysize=5cm
\centerline{\epsfbox{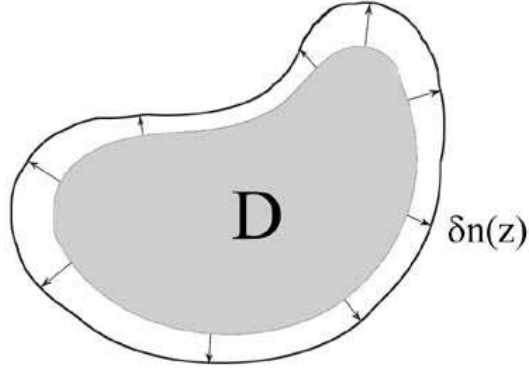}}
\caption{\sl The domain $D$ and vectors of normal
displacement.}
\label{fi:deltan}
\end{figure}

A change of the boundary can be characterized
          by its normal displacement $\delta n (z)$, such
          that $\delta n (z)$ is a continuous
function on the boundary (see Fig.~\ref{fi:deltan}).
The Hadamard formula \cite{Hadamard} expresses the deformation of the
Green function through the Green function itself:
\beq\label{Hadamard}
\delta G(z_1,z_2)=\frac{1}{2\pi}\oint_{\p D}\p_nG(z_1,\xi)\p_nG(z_2,\xi)
\,\delta n(\xi)|d\xi|.
\eeq
A corollary of the Hadamard formula is the variation of the boundary
value of a harmonic function under variation of the boundary. It reads
\beq\label{three1}
\delta f^H (z) =({\sf R} f)(z)\,\delta n(z),\quad z\in\p D.
\eeq
\end{itemize}

\section{Large $N$ limit}

The large $N$ limit  is understood as
$N \to \infty$, $\hbar \to 0$ while  $\hbar N$ is kept finite.
The expansion
in $N^{-1}$ is then equivalent to the expansion in $\hbar$.
We call it a semiclassical limit.

\paragraph{Semiclassical density and the support of the eigenvalues.}
To elaborate the semiclassical limit, we have to find the maximum of the
integrand $ |\Psi_N(z_1,\dots,z_N)|^2$
in  (\ref{psi}). At finite $N$,
it is given by the conditions
$\p {\log|\Psi_N|}/\p z_i = 0$ for every $i$:
$$
\hbar \!\sum_{j=1, \neq i}^{N}\frac{1}{z_i -z_j}
+\p_{z_i}W(z_i) =0,
$$
or by the equation
\beq\label{semitau1}
\p_z\left(\varphi_{0}(z)+ W(z)\right)=
\p_{\bar z}\left(\varphi_{0}(z)+ W(z)\right) =0\,,
\quad \quad z\in D
\eeq
in the limit.
Here $\varphi_{0}(z)=
\left <\varphi (z)\right >=
\int \log|z-z'|^2\left<\rho(z)\right> (z')d^2z' $ and
$\left<\rho(z)\right>$
          are expectation values
of the potential and the density in the leading semiclassical
approximation.
The last equation holds inside the support of
eigenvalues, $D$. It does not hold outside it.
In the semiclassical approximation, $D$ is a
domain with a well-defined sharp boundary
determined by the potential $W$. An assumption that the support
is a compact
connected domain bounded by a Jordan curve implies
       restrictions on the potential.
The restrictions do not reduce
       the number of  parameters in the potential but rather
ranges of their variation.
We do not discuss them here. The assumption is valid,
for example, for small
perturbations of the Gaussian potential.

The  solution for density is obtained by applying $\p_{\bar z}$
to eq.\,(\ref{semitau1}):
\beq\label{rho0}
\left<\rho(z)\right> =\left \{
\begin{array}{ll}
\sigma (z) &\quad \mbox{if}\quad z\in D
\\&\\
0 &\quad \mbox{ if}\quad z \in{\bf C}\setminus D.
\end{array}
\right.
\eeq
The function $\sigma$ is introduced
in (\ref{sigma}).
Consequently,
$\displaystyle{\varphi_{0}(z)=}\int_D \log|z-z'|^2
\sigma (z')d^2z'$.
This function is harmonic in the exterior of the domain
except at infinity where it has a logarithmic singularity.
Eq. (\ref{semitau1})
means that inside $D$, including the boundary,  it is equal to $-W$
plus a constant. Since $W(0)=0$, the constant is
$\displaystyle{\varphi_{0}(0)=
\int_D \log|\xi |^2\sigma (\xi ) d^2\xi}$, and so
$\varphi_{0}(z)+W(z)=\varphi_0 (0)$ for $z\in D$.
Moreover, according to (\ref{semitau1}),
\beq\label{prop}
\p_n\left(\varphi_{0}(z)+W(z)\right)=0,\quad \quad z\in\p D,
\eeq
so  both tangential and normal derivatives
of $\varphi_0 +W$ at the boundary vanish.

Since
$\varphi_0(z)-N\hbar \log |z|^2$ is harmonic outside,
the harmonic extension of $W$ is
\beq\label{extW}
W^H(z)=\varphi_0(0)-\varphi_0 (z)+
N\hbar \left ( \log |z|^2 -\left (\log |z|^2 \right )^H \right ),
\eeq
whence it follows that
\beq
{\sf R}\left ( W(z)+N\hbar \log |z|^2 \right )=0\,.
\eeq
This condition means, in other words,
that the domain $D$ is
such that the function $\p_z W$ on its boundary is the
boundary value of an analytic function in
${\bf C}\setminus D$.
The latter, together with  the normalization
condition
$\int_D\Delta W d^2z=-4\pi\hbar N$,
determines the shape of the support of eigenvalues.

For the potential of the form (\ref{potential})
with $V(z)=\sum_{k\geq 1}t_k z^k$
this condition is somewhat more explicit.
In this case, the shape
of the domain is determined,  by the relations
$$
-\frac{1}{\pi k}
\int_{ D}z^{-k}dz =t_k\,,
\;\;\;\;
\frac{1}{\pi}\int_{D} d^2z =\hbar N.
$$
such that$\pi \hbar N$ is the area
of $D$ and $-\pi k t_k$ are harmonic moments of the
domain complementary to $D$. The problem is thus
equivalent to the inverse problem of 2D potential
theory.

Using the relation $W(z)=\varphi_{0}(0)- \varphi_0 (z)$,
it is easy to find the value of
$\hbar^2 \log|\Psi_N|^2$ at the saddle point, which
we denote by $F_0$:
\beq\label{semitau2}
F_0
=-\, \int_D \!\int_D \log \left |
\frac{1}{z} -\frac{1}{z'}\right |
\sigma (z) \sigma (z')d^2z d^2z'.
\eeq
The leading asymptote of the partition function as $\hbar \to 0$
is therefore $Z_N \simeq e^{F_0/\hbar^2}$.
By  inspection one can check that the average density (\ref{rho0}) can be
also obtained by variation of  (\ref{semitau2}): $\left < \rho (z)\right
>=\delta F_0 /\delta W(z)$.

      The function $F_0$ plays an important role in
the complex analysis. It generates conformal maps from
the exterior of $D$ onto the unit disk and gives a
formal solution to the Dirichlet boundary problem.
See Sec. 4 of \cite{MWZ} for
details.

\paragraph{Variation of the support of eigenvalues.}
Let us examine the change of the support of eigenvalues $D$
under a small change of the potential at fixed $N$.
In general, we can write: $\int_{\delta D}f(z)d^2z =
\oint_{\p D} \delta n(z)f(z)|dz|$ for any function $f$,
where $\delta D$ stands for a  strip between the domains
$D (W+\delta W)$ and  $D(W)$ and $\delta n$ has the same
meaning as in the previous section (Fig.~\ref{fi:deltan}).

The variation of the saddle point condition is most
conveniently found from (\ref{prop}):
$$
\delta(\p_n(W+\varphi_{0}))=\delta n \, \p_n^2(W+\varphi_{0})+
\p_n\delta(W+\varphi_{0})=0.
$$
where $z$ is on the boundary.
Since both tangential and normal derivatives of
$W+\varphi_{0}$ vanish on the
boundary, one writes
$\p_n^2(W+\varphi_{0})=
\Delta  (W+\varphi_{0})=\Delta W=-4\pi\sigma$.
This gives the variation of the boundary:
\beq\label{CC4}
\delta n(z) =\frac{1}{4\pi\sigma (z)}({\sf R}\,\delta W).
\eeq
          Here ${\sf R}\,\delta W
=\p_n (\delta W -(\delta W)^H)$ is the external Neumann
jump operator defined by (\ref{R}). (We used the fact that
$-\delta \varphi_{0}$ is, up to a constant,
the harmonic extension of $\delta W$ to the exterior domain,
as is seen from (\ref{extW}).)
Note that if the variation of the potential is harmonic outside
$D$ including
infinity the domain does not change.

Let us check that our result meets the requirement
that $N$ stays constant under a variation of the potential. Since
$$
\oint \delta n(z)\sigma (z)|dz|=
\frac{1}{4\pi}\oint \p_n \delta W(z) |dz| =
\frac{1}{4\pi}\int_D \Delta \delta W (z) d^2z
$$
it is easy to see that
$$
\delta \int \left < \rho (z) \right >d^2z
= \oint \delta n(z)\sigma (z)|dz|
+ \int_D \delta \sigma (z) d^2z =0,
$$
so that  $N$ is kept constant under this variation.

Consider special variations of the potential of the form $\delta_\zeta
W=\varepsilon\log|z-\zeta|^2$. Then one is able to express
${\sf R}\delta_{\zeta}W$ through the Dirichlet Green function
(cf. (\ref{log})).
It is convenient to introduce the modified Green function:
\beq
\label{green}
{\cal G}(z,\zeta)=G(z,\zeta)-G(z,\infty)-G(\infty,\zeta)\,.
\eeq
It is easy to see that
$$
\delta_{\zeta}
n(z)=\frac{1}{2\pi\sigma(z)}\p_n{\cal G}(z,\zeta ).
$$
(The last term in ${\cal G}$
       vanishes under
the normal derivative. It is  included
       for the symmetry and future convenience.)

\paragraph{Variation of the boundary under variation of the size of
the matrix.} If one varies the size of the matrix, $N \to N+\delta N$,
keeping the potential fixed, the support of
eigenvalues also changes its shape. To find how it grows,
we note that $\delta W=0$
means that $\delta \varphi_0(z)=\delta \varphi_0(0)$
for all $z\in D$. Plugging here the integral representation
of $\varphi_0$, we conclude that
$\oint \log |z^{-1}-\xi^{-1}|\delta n(\xi)\,\sigma (\xi)
|d\xi| =0$ for all $z$ in $D$, with the variation
of the normalization condition being
$\hbar\delta N = \oint \delta n(z)\sigma (z)|dz|$.
These conditions are met with
\beq\label{var1}
\delta n(z) =-\frac{\hbar \delta N}{2\pi\sigma (z)}
\p_n G(z,\infty),
\eeq
where $G(z,\infty)=-\log|w(z)|$.
This describes the interface dynamics in Laplacian growth models,
known as Darcy's law (cf.\,\cite{MWWZ}).

\section{Connected correlation functions in the
first non-vanishing order in $\hbar$.}

\paragraph{Connected two-point function.} In order to obtain the
two-point function one has to vary the one point function
$\left <\rho (z)\right > =
\Theta (z; D)\,\sigma (z)$, where $\Theta (z;D)$ is the characteristic
function of the domain $D$:
$\Theta (z;D)=1$ for $z\in D$ and
$\Theta (z;D)=0$ for $z\notin D$. The variation
reads
\beq\label{CC}
\delta \left <\rho \right >
=\delta \sigma \,\Theta (z; D)
+\sigma \,\delta \Theta (z; D).
\eeq
The second term is localized on the contour.
Let $\delta(z;\p D)$ be a $\delta$-function located on the boundary
of $D$, defined by the condition
that $\int f(z)\delta(z;\p D)d^2z=\oint_{\p
D}f(z)|dz|$ for any smooth function $f$.
It is clear that
$\delta \Theta (z;D) =\delta n(z) \delta (z;\p D)$.
Using the relation  (\ref{CC4}) between a variation of the potential and
deformation of the domain, we write:
$$
4\pi  \delta \left < \rho  \right >
=-\Delta \delta W  \, \Theta (z;D)
+({\sf R}\,\delta W ) \,\delta (z;\p D ).
$$

To keep track of the singular boundary terms,
it is helpful to integrate the variation of density
with some reasonable
function $f$ on the plane. This is equivalent
to calculating $\delta \left <\mbox{tr}\, f(M)\right >$
instead of $\delta \left <\rho \right >$.
Setting $\delta W=\varepsilon g$
we have:
$$
\frac{4\pi}{\varepsilon}
\int \delta \left < \rho  \right > f d^2z =
- \int_D \Delta g\, f d^2z  +
\oint_{\p D} f ({\sf R} g)\,|dz|\,.
$$
The result is symmetric with respect to
$f\leftrightarrow g$. With a help of
the Green formula, it  can be
expressed through the Bergmann kernel (\ref{CC51}):
\beq\label{CC61}
\begin{array}{c}
\hbar^{-2}\displaystyle{\left <\mbox{tr}\, f(M)\, \mbox{tr}\,
g(M)\right >_{{\rm conn}}}
= \displaystyle{\frac{1}{\pi}
{\cal R}e \int_D \p_z f (z) \p_{\bar z}g(z) d^2z \, \, +}
\\ \\
+\,\displaystyle{\frac{1}{8\pi^2}
\oint \! \oint f(z)\p_n \p_{n'}G(z,z')g(z') |dz||dz'|}.
\end{array}
\eeq

Choosing $f(z)=\log |z_1 -z|^2$,
$g(z)=\log |z_2 -z|^2$,
we find the
pair correlation function
of the Bose field $\varphi(z)$:
\beq\label{phi}
\frac{1}{2\hbar^2}\left < \varphi(z_1)\varphi(z_2)
\right >_{{\rm conn}}
\,=
      {\cal G}(z_1 , z_2 ) -\log \frac{|z_1 -z_2|}{r},
\eeq
where ${\cal G}$ is introduced in (\ref{green})
and $\log r =\lim_{z\to \infty}(\log |z |
+G(z, \infty ))$ is Robin's constant ($r$ is the exterior
conformal radius of the domain $D$, see e.g. \cite{Hille}).
The function in the r.h.s. is harmonic outside the domain.
If one of the points, say $z_1$, is located inside,
one sets the corresponding Green functions in (\ref{green})
to be zero. In particular, if both points are inside,
the correlation
function is just
\beq\label{CC81}
\frac{1}{2\hbar^2}\left < \varphi(z_1)\varphi(z_2) \right >_{{\rm conn}}
\,=- \, \log \frac{|z_1-z_2|}{r}.
\eeq
This result is valid for well separated points
($|z_1-z_2|^2>>N\hbar/\sigma$).

Taking holomorphic
or antiholomorphic derivatives of (\ref{phi}),
we find pair correlations of currents:
\beq\label{CC9}
\hbar^{-2}\left <J(z_1)J(z_2)\right >_{{\rm conn}}
= - \frac{1}{(z_1 - z_2 )^2} + 2\p_{z_1}\p_{z_2}G(z_1 ,z_2 ),
\eeq
\beq\label{CC10}
\hbar^{-2}\left < J(z_1)\, \bar J(z_2)\right >_{{\rm conn}}
= - \pi \delta^{(2)}(z_1-z_2 ) +2
\p_{z_1}\p_{\bar z_2}G(z_1 ,z_2 )
\eeq
(here it is implied that both points are outside).
These results resemble the two-point functions
of the Hermitian 2-matrix model found in \cite{DKK}; they were
also obtained in \cite{J} in the study of thermal fluctuations of a
confined 2D Coulomb gas.

Outside the domain formulas (\ref{phi}), (\ref{CC9})
describe
correlations also at merging points away from the boundary. In
particular, the mean square fluctuation of the current is
\beq\label{CC11}
\hbar^{-2}\left <  J^2(z) \right >_{{\rm conn}}
=\frac{1}{6}\{w;z\}\,,
\;\;\;\;\;z\in {\bf C}\setminus D.
\eeq
Here
\beq\label{CC12}\{w;z\}=
\frac{w'''(z)}{w'(z)} -\frac{3}{2}
\left ( \frac{w''(z)}{w'(z)}\right )^2 =
6\, \lim_{z'\to z}
\left ( 2\p_z \p_{z'} G(z,z') -
\frac{1}{(z-z')^2}\right )
\eeq
is the Schwarzian derivative
of the conformal map $w(z)$.

These formulas show that there are local correlations
in the bulk as well as strong long range correlations
at the edge.
(See \cite{Jancovici82} for a similar result in the context
of classical Coulomb systems).
At the same time
further variation of the pair density correlation
function suggests that, starting from $n =3$, the connected
$n$-point density correlations
vanish in the bulk in all orders of $\hbar$
(in fact they are exponential in $1/\hbar$).
All the leading contribution (of order $\hbar^{2n-2}$)
comes from the boundary.

Note that the result (\ref{CC61}) is universal in the sense
that it depends on the potential only through the
form of the domain $D$. The universality holds
for any connected correlation function
of two traces.

The kernel in the boundary term in (\ref{CC61}) is
the absolute value of the Bergman
kernel \cite{Bergman}
of the domain ${\bf C}\setminus D$ at the boundary.
Presumably, this result can be generalized to
the more complicated case of non-connected
supports of eigenvalues, with the boundary
term being expressed through the Bergman
kernel of the Schottky double of the Riemann
surface ${\bf C}\setminus D$.
For the Hermitian one-matrix model, where the
support of eigenvalues shrinks to a number of
cuts on the real axis, a similar result was
recently obtained in \cite{CG}.

\paragraph{Connected three-point function.}
The 3-point function can be obtained by further
varying eq.\,(\ref{CC61}). Let us first transform
the r.h.s., to bring it to the form convenient
for the varying. Using the Green formula, one rewrites the two point
function as an integral over the entire complex plane plus the
integral over the exterior of the domain
$$
4\pi \hbar^{-2}\left < \mbox{tr}\, f(M)\, \mbox{tr}\,
g(M)\right >_{{\rm conn}}=-
\int_{{\bf C}} f\Delta gd^2z +
\int_{{\bf C}\setminus D} (g-g^H ) \Delta f d^2z\,.
$$
The variation of the first term
is zero since it does not depend on the contour.
The variation of the second term consists of two parts:
the one coming from variation of the boundary and another
one from variation of the integrand. The first part
actually vanishes because the integrand equals zero
on the boundary, where $g=g^H$. The variation of the integrand  is
$
\displaystyle{
-\int_{{\bf C}\setminus D}(\delta g^H )\Delta f \, d^2z =
\oint (\delta g^H ) ({\sf R} f)
|dz|}
$
(where we substituted $f\to f-f^H$ under the Laplace operator
and again  used the Green formula).
Next, we use the Hadamard formula (\ref{Hadamard})
to compute the $\delta g^H$,
the response of the harmonic extension
of the function $g$ to a small
change of the domain, taken on the boundary of the
initial domain (\ref{three1}). The result is
$$
4\pi \hbar^{-2}\delta  \left < \mbox{tr}\, f(M)\, \mbox{tr}\,
g(M)\right >_{{\rm conn}}=
\oint ({\sf R}\,f)({\sf R}\,g)\,
\delta n |dz|.
$$
Now, let us redefine $f=f_1$, $g=f_2$ and
plug (\ref{CC4}) for the $\delta n(z)$ with
a function $f_3$: $\displaystyle{\delta n =
\frac{\varepsilon}{4\pi\sigma}{\sf R}f_{3}}$.
Using (\ref{FD6}), we
obtain the correlation function of three traces:
\beq\label{three2}
\left <
\prod_{j=1}^{3}\mbox{tr}\, f_j (M)\right >_{{\rm conn}} =
\frac{\hbar^{4}}{16\pi^2 }\oint \frac{|dz|}{\sigma (z)}
\prod_{j=1}^{3}\, {\sf R} f_i (z).
\eeq
The answer is non-universal, i.e., it depends explicitly
on the boundary value of the Laplacian of the potential.
Note that if at least one of the functions $f_j$ is
harmonic in ${\bf C}\setminus D$, then the correlation function vanishes
(in this leading order in $\hbar$ of course).

Alternatively, one can apply the Hadamard formula directly to the
2-point correlation function
of the potentials $\varphi$ (\ref{phi}). We have
$$ \left <
\varphi(z_1) \varphi(z_2)\varphi(z_3)\right>_{{\rm conn}}
=\frac{\hbar^2}{\varepsilon}
\delta_{z_3}\left <
\varphi(z_1) \varphi(z_2)\right>_{{\rm conn}}=2
\frac{\hbar^4}{\varepsilon} \delta_{z_3}
\left [{\cal
G}(z_1,z_2) +\log r \right ]\, =$$
$$=\frac{\hbar^4}{\varepsilon \pi}
\oint \p_n{\cal G}(z_1,\xi)\p_n{\cal
G}(z_2,\xi)\,\delta_{z_3}n(\xi)\,|d\xi|\, =$$
$$=\frac{\hbar^4}{2\pi^2}
\oint \frac{|d\xi|}{\sigma(\xi)} \p_n{\cal G}(z_1,\xi)\p_n{\cal
G}(z_2,\xi)\,\p_n{\cal G}(z_3,\xi)\,.$$
The result agrees with the formula for the third
order derivatives of the tau-function obtained in
\cite{MWZ} within a different approach (also based on the
Hadamard variational formula).

For example, in the case $W=-|z|^2$, the support is a
disk of the radius $R=\sqrt{\hbar N}$.
The conformal map is simply $w(z)=z/R$
and the above formula gives
$$
\left <\prod_{j=1}^{3}\varphi (\lambda_j)
\right >_{{\rm conn}}=-2\hbar R^2 {\cal R}e \, \left (
\frac{1}{(\lambda_1 \bar \lambda_2 -R^2 )
(\lambda_1 \bar \lambda_3 -R^2 )}
+\, [1\leftrightarrow 2 ],\ +\,
[1\leftrightarrow 3 ]\, \right ).
$$
Finite $N$ formulas for the general $n$-point correlations reviewed
in \cite{F} should be able to be used to reclaim the same result.

\paragraph{Variation of contour integrals.}
One may proceed in the same way to find
the connected $n$-point correlation functions.
Starting from $n=4$, however, one encounters  technical
difficulties.

When passing
from 2-point to 3-point functions we transformed
      integral over the boundary into a bulk integral since the
latter is easier to vary. Passing to  4-point functions,
one needs to vary
the contour integral in (\ref{three2}) which does not seem
to be naturally representable in a bulk form.
We have to learn how to vary contour integrals.
Here are general rules.

Consider the contour integral of the general form
$\oint F(f (z), \p_n f(z))ds$ where $ds =|dz|$ is the line
element along the boundary curve and
$F$ is any fixed function. Calculating the linear response
to the deformation of the contour,
one should vary all items in the integral independently
and add the results. There are four elements to be varied:
the support of the integral $\oint$, the $\p_n$, the line element $ds$
and the function $f$. By variation of the $\oint$ we mean
integration of the old function over the new contour. This
gives
$\oint \delta n \, \p_n F \,ds.$
The change of the slope of the normal vector results in
$\delta \frac{\p}{\p n} =-\p_s (\delta n) \frac{\p}{\p s}.$
The rescaling of the line element gives
$\delta ds =\kappa \,\delta n ds$,
where $\kappa (z)$ is the local curvature of the boundary curve.
The  curvature is
$\kappa =d\theta /ds$
where $\theta$ is the angle between the outward pointing
normal vector
to the curve and the $x$-axis.
The formula
$\kappa (z)=\p_n \log \left |w(z)/w'(z)\right |$
expresses the curvature through the conformal map. This formula is
useful in calculating the 5-point function. We do not attempt to do
this here. Another element is
the Laplace operator on the boundary in terms of normal ($\p_n$)
and tangential ($\p_s$) derivatives:
$\Delta = \p_n^2 +\p_s^2 +\kappa \p_n$.

At last, we have to vary the function $f$
if it explicitly depends on the contour.
In particular, if this function is the harmonic
extension of a contour-independent function on the
plane, its variation on the boundary is given by
(\ref{three1}).

\paragraph{Connected four-point function.}
Let us elaborate the case $n=4$.
In accordance with (\ref{FD6}),
we find
$\left <
\prod_{j=1}^{4}\mbox{tr}\, f_j (M)\right >_{{\rm conn}}$
from the response of the r.h.s. of eq.\,(\ref{three2})
to the variation of the potential $\delta W =\varepsilon f_4$,
so that
\beq\label{four0}
\delta \sigma =-\frac{\varepsilon}{4\pi }\Delta f_4\,,
\;\;\;\;\;\;
\delta n =\frac{\varepsilon}{4\pi \sigma}
{\sf R}f_4.
\eeq

To get the result, we apply the above rules to the contour integral
(\ref{three2}).
Note that in this particular case a change of the slope of the normal
vector gives no contribution since the functions under the normal
derivative do not change along  the boundary. Along this way we obtain the
following result:
\beq\label{four2}
\begin{array}{c}
\displaystyle{
\frac{64 \pi^3}{\hbar^6}
\left < \prod_{j=1}^{4} \mbox{tr}\, f_j (M) \right >_{{\rm conn}}
=\oint \frac{|dz|}{\sigma^2 }
\sum_{i=1}^{4}\Delta f_i  \prod_{k=1, \neq i}^{4}
{\sf R}f_k\, -}
\\ \\
-\,
\displaystyle{
\oint \frac{|dz|}{\sigma^2 }
(\p_n \log \sigma  +2\kappa  )\prod_{k=1}^{4}
{\sf R}f_k\, -}
\\ \\
-\,
\displaystyle{
\oint |dz|  \left [
\frac{{\sf R}f_1 {\sf R}f_2}{\sigma  }\,
\p_n \! \left (\frac{{\sf R}f_3
{\sf R}f_4}{\sigma }\right )^H \, +
[ 1\leftrightarrow 3] \, + \, [ 2\leftrightarrow 3] \,\right ]}.
\end{array}
\eeq
The first two terms are explicitly symmetric with
respect to all permutations of
$(1,2,3,4)$. The third
one is seemingly not, but in fact it is, as is clear from the Green
formula.

In the 4-point correlation function
the first term in the r.h.s.
of (\ref{four2}) vanishes.
For $\sigma (z)=1/\pi$, we get
\beq\label{four21}
\begin{array}{c}
\displaystyle{
\left < \prod_{j=1}^{4}
\varphi (\lambda_j)
\right >_{{\rm conn}}  =
\frac{\hbar^6}{8\pi^2}\oint \! |dz| \!\,\p_{n} \oint \! |dz'|
\, \p_{n'}{\cal G}(z,z')\, \times }
\\ \\
\times \,\, \displaystyle{
\Bigl (\p_{n}{\cal G}(\lambda_1 , z)\p_{n}
{\cal G}(\lambda_2 , z)
\p_{n'}{\cal G}(\lambda_3 , z')\p_{n'}{\cal G}(\lambda_4 , z')
+ [1\leftrightarrow 3] + [2\leftrightarrow 3] \Bigr )} \, -
\\ \\
\displaystyle{-\frac{\hbar^6}{2\pi}\oint |dz| \kappa(z)
\prod_{j=1}^{4}\p_{n}{\cal G}(\lambda_j , z),}
\end{array}
\eeq
where ${\cal G}$ is defined by (\ref{green}) and
$\lambda_j$'s are assumed to be outside.

\section{Discussion}

We have shown that large scale properties of the normal matrix ensemble
are obtained from the analytical properties of a curve on the complex
plane.  The
curve bounds  a semiclassical support of eigenvalues and is
determined by the relation $ \p W(z)=-\hbar
\left <\mbox{ tr}
\frac{1}{z-M}\right >$.

Large scale correlation functions of the ensemble are objects
of the Dirichlet boundary problem
for the non-compact exterior domain
complementary to the compact domain $D$.
The two point function is essentially the Dirichlet Green
function, the higher order functions are related  to the deformation
of the Green function under deformations of the curve. They are
determined by successive  applications of the Hadamard
formula and are therefore
expressed through the Neumann jump on the curve and
through the Bergman kernel.

We expect that other objects of  matrix ensembles, such as
the genus ($1/N$)
expansion  of the partition function  are recorded in the analytic
properties of the curve. In particular, we expect that  $F_1$ -  a genus 1
correction to the partition function $\log Z_N$ (a correction of
order $\hbar^0$) is related to the determinant
of the Laplace operator in the exterior domain
${\bf C}\setminus D$.

\section*{Acknowledgments}
We acknowledge  dis\-cus\-sions with
O.Agam, J.Amb\-jorn, E.Bet\-tel\-heim,
O.Bo\-hi\-gas, A.Bo\-yar\-sky, A.Ca\-ce\-res,
L.Che\-khov, A. Gor\-sky,
V.Ka\-za\-kov, I.Kos\-tov, Y.Ma\-ke\-en\-ko,
A.Mar\-sha\-kov, M.Mi\-ne\-ev-\-Wein\-stein,
O.Ru\-chay\-skiy,  R.Theo\-do\-res\-cu and
P.Zinn-\-Jus\-tin. We are indebted to P.J.For\-res\-ter
who drew our attention
to Refs. \cite{J,F}.
A.Z. thanks B.Jancovici for useful remarks. P.W.
was supported by grants NSF DMR 9971332 and MRSEC NSF DMR 9808595. P.W.
thanks S.Ouvry for the hospitality in LPTMS  at Universite de Paris Sud
at Orsay, where the paper was completed.
The work of A.Z. was supported in part by
RFBR grant 01-01-00539,  by grant INTAS-99-0590 and by
grant 00-15-96557 for support of scientific schools.

%%%%%%%%%%%%%%%%%%%%%%%%%%%%%%%%%%%%%%%%%%%%%%%%%%%%%%%%%%%%%%%%%%%%%%


\begin{thebibliography}{66}
\bibitem{Ginibre}J.Ginibre, J. Math. Phys. {\bf 6} (1965) 440,
V.Girko Theor. Prob. Appl. {\bf 29} (1985), 694;
ibid  {\bf 30} (1986) 677

\bibitem{DKK} J.-M.Daul, V.Kazakov and I.Kostov,
Nucl. Phys. {\bf B409} (1993) 311-338

\bibitem{c1} S.Alexandrov, V.Kazakov and I.Kostov,
to appear in Nucl. Phys. B,
e-print archive: hep-th/0205079


\bibitem{MQM}B.Eynard, J. Phys. A {\bf 31} (1998) 8081,
e-print archive: cond-mat/9801075

\bibitem{recent}Y.Fyodorov, B.Khoruzhenko and
H.-J.Sommers, Phys. Rev. Lett. {\bf 79} (1997) 557,
e-print archive: cond-mat/9703152;
J.Feinberg and A.Zee, Nucl. Phys. {\bf B504} (1997) 579-608,
e-print archive: cond-mat/9703087;
G.Akemann, e-print archive: hep-th/0204246



\bibitem{MWWZ}
M.Mineev-Weinstein, P.B.Wiegmann and A.Zabrodin,
Phys. Rev. Lett. {\bf 84} (2000) 5106-5109,
e-print archive: nlin.SI/0001007

\bibitem{WZ}P.B.Wiegmann and A.Zabrodin,
Commun. Math. Phys. {\bf 213} (2000) 523-538,
e-print archive: hep-th/9909147;

\bibitem{KKMWZ}
I.Kostov, I.Krichever,
M.Mineev-Weinstein, P.Wiegmann and A.Zabrodin, {\it $\tau$-function for
analytic curves}, Random matrices and their applications, MSRI
publications, eds. P.Bleher and A.Its,
vol.40, p. 285-299, Cambridge Academic Press, 2001,
e-print archive: hep-th/0005259;

\bibitem{Z} A.Zabrodin,
Teor. Mat. Fiz. {\bf 129} (2001) 239-257 (in Russian,
English translation:
Theor. Math. Phys. {\bf 129} (2001) 1511-1525),
e-print archive: math.CV/0104169

\bibitem{MWZ} A.Marshakov, P.Wiegmann and A.Zabrodin,
Commun. Math. Phys. {\bf 227} (2002) 131-153,
e-print archive: hep-th/0109048

\bibitem{ABWZ} O.Agam, E.Bettelheim, P.Wiegmann and A.Zabrodin,
Phys. Rev. Lett. {\bf 88} (2002) 236801,
e-print archive: cond-mat/0111333



\bibitem{ChauYu} L.-L.Chau and Y.Yu,
Phys. Lett. {\bf 167A} (1992) 452

\bibitem{ZC} L.-L.Chau and O.Zaboronsky,
Commun. Math. Phys. {\bf 196} (1998) 203-247,
e-print archive: hep-th/9711091


\bibitem{J}A.Alastuey and B.Jancovici,
J. Stat. Phys. {\bf 34} (1984) 557; B.Jancovici
J. Stat. Phys. {\bf 80} (1995) 445

\bibitem{F}P.J.Forrester, Physics
     Reports, {\bf 301} (1998) 235-270


\bibitem{univ1} J.Ambjorn, J.Jurkiewicz
and Y.Makeenko, Phys. Lett. {\bf B251} (1990) 517;
J.Ambjorn, L.Chekhov, C.F.Kristjansen and Yu.Makeenko,
Nucl. Phys. {\bf B404} (1993) 127-172; Erratum: ibid.
{\bf B449} (1995) 681, e-print archive: hep-th/9302014

\bibitem{univ2}E.Brezin and A.Zee, Nucl. Phys. {\bf B402} (1993) 613

\bibitem{int} H.Aratyn, Lectures presented
at the VIII J.A. Swieca Summer School,
Section: Particles and Fields, Rio de Janeiro - Brasil - February/95,
e-print archive: hep-th/9503211;
M.Adler and P. van Moerbeke,
Ann. of Math. (2) {\bf 149} (1999), no. 3, 921-976,
e-print archive: hep-th/9907213


\bibitem{Mehta} M.L.Mehta, {\it Random matrices},
Academic Press, NY, 1967



\bibitem{fermions} T.Banks, M.Douglas, N.Seiberg and
S.Shenker, Phys. Lett. {\bf B238} (1990) 279

%\bibitem{FJ1996} P.J.Forrester and B.Jancovici,
%Int. J. Mod. Phys. {\bf A11} (1996) 941-949

\bibitem{FGIL} P.Di Francesco, M.Gaudin, C.Itzykson and
F.Lesage, Int. J. Mod. Phys. {\bf A9} (1994) 4257-4351

\bibitem{Jev}S.Das and A.Jevicki, Mod. Phys. Lett.
{\bf A5}(1990) 1639

\bibitem{C-H} A.~Hurwitz and R.~Courant,
{\it Vorlesungen \"uber allgemeine Funktionentheorie und
elliptische Funktionen. Herausgegeben und erg\"anzt durch
einen Abschnitt \"uber geometrische Funktionentheorie},
Springer-Verlag, 1964 (Russian translation, adapted
by M.A.~Evgrafov: {\it Theory of functions}, Nauka, Moscow,
1968)

\bibitem{Gakhov} G.Gakhov {\it Boundary value problems}, Dover,
NY, 1990

\bibitem{Hadamard}J.Hadamard, Mem. presentes par divers savants a
l'Acad. sci., {\bf 33} (1908).

\bibitem{Hille} E.Hille, {\it Analytic function theory},
v.II, Ginn and Company, 1962



\bibitem{Jancovici82} B.Jancovici, J. Stat. Phys. {\bf 28}
(1982) 43

\bibitem{Bergman} S.Bergman, {\it The kernel function and
conformal mapping}, Math. Surveys {\bf 5}, AMS, Providence, 1950

\bibitem{CG} Y.Chen and T.Grava, to appear in J. Phys. A,
e-print archive: math-ph/0201024




\end{thebibliography}
\end{document}